\documentclass[10 pt,aps,twocolumn,preprint,showpacs,preprintnumbers,amsmath,amssymb]{revtex4}

\usepackage{graphicx}

\begin{document}

\title{Single-electron charging and detection in a laterally-coupled quantum 
dot circuit in the few-electron regime}

\author{L.-X. Zhang, P. Matagne and J.P. Leburton
}
\affiliation{
Beckman Institute for Advanced Science \& Technology\\
and Department of Electrical and Computer Engineering, \\
University of Illinois at Urbana-Champaign, Urbana, Illinois 61801\\
}
\author{
R. Hanson and L. P. Kouwenhoven
}
\affiliation{
Department of NanoScience and ERATO Mesoscopic Correlation Project, \\
Delft University of Technology, \\
P.O. Box 5046, 2600 GA Delft, The Netherlands} 
\date{\today}


\begin{abstract}

We provide a physical analysis of the charging and detection of the 
first few electrons in a laterally-coupled GaAs/AlGaAs quantum dot 
(LCQD) circuit with integrated quantum point contact (QPC) read-out. Our 
analysis is based on the numerical solution of the Kohn-Sham 
equation incorporated into a three-dimensional self-consistent 
scheme for simulating the quantum device. Electronic states and 
eigenenergy spectra reflecting the particular LCQD confinement shape are 
obtained as a function of external gate voltages. We also derive the 
stability diagram for the first few electrons in the device, and 
obtain excellent agreement with experimental data. 

\end{abstract}

\maketitle

\section{Introduction}
\label{sec:intro}

Lateral GaAs/AlGaAs quantum dots (QD's) are now routinely fabricated 
with planar technology.\cite{Kastner} Three-dimensional (3D) quantum 
confinement is achieved, in part, by using the GaAs/AlGaAs 
semiconductor heterostructures to confine the conduction electrons 
into a two dimensional electron gas (2DEG) at the interface between 
the two materials. By placing metal gates on top of such a structure, 
carrier confinement in other in-plane directions can be realized 
by energizing the gates that create lateral energy barriers 
to electrons in the 2DEG. Design of these QD's, which previously 
contained tens of electrons, has been improved to operate them in 
a few-electron regime where the charging of the very first electrons 
can be observed experimentally.\cite{Ciorga} Two quantum dots can be 
placed adjacent to each other to form a laterally coupled device 
with both electrostatic and quantum-mechanical coupling between 
them.\cite{Waugh, Elzerman} Fine variations of the top gate biases change 
the confinement of each dot, while precise coupling between them through 
the central gates leads to a fully tunable two-qubit quantum system, 
which can be used as a building block for quantum computing.\cite{Loss}

Recently, it has been shown that laterally-coupled quantum dots 
(LCQD) containing a few conduction electrons could be coupled to 
single charge detectors to form an integrated quantum circuit.\cite{Elzerman} 
The read-out of the charge state in the LCQD is 
realized by integrating monolithically quantum point contacts 
(QPC's) adjacent to each of the QD's. Each QPC can be calibrated through 
electrostatic coupling with the dots so that its conductivity changes 
abruptly once a single-electron  charging event occurs in one of the dots.\cite{Field, Sprinzak} 
With this sensitive detector, it is then possible 
to obtain the ``stability diagram'' that describes the stable charge 
regimes of the LCQD as a function of the tuning (plunger) gate biases.\cite{Pothier,Wiel}

This quantum dot circuit has a two-fold advantage: it is possible (i) to 
scale it to a quantum dot array, (ii) to perform single-quantum 
sensitivity measurements, both of which are favorable features of a 
realizable quantum computer.\cite{Bennett}

In this paper we study the properties of the above circuit via 
numerical simulation that involves the self-consistent solution of 
coupled Poisson and Kohn-Sham equations discretized on a 3D mesh.\cite{Mat1,Mat2} 
In Sec. II we describe the 
LCQD structure and in Sec. III we present the approach for solving the 
Kohn-Sham equations in the device environment within the local spin 
density approximation (LSDA) and express the criterion used to determine
the charging events as a function of the applied gate biases. In Sec. IV 
we present our simulation results of the circuit, including both 
electrostatic and quantum-mechanical features, the functionality of the 
QPC's, and the stability diagram in the few-electron charging regime. 
Finally, we summarize our work in Sec. V.

\section{Dot Structures}

Figure 1(a) shows the top view of the LCQD and QPC gates in the $xy$-plane.\cite{Elzerman} 
Top L-, R-, T- and M-gates are used to define the two 
coupled-dot region. Among them, the T- and M-gates can also control the 
coupling between the two dots. The PL- and PR-gates, called the ``plungers,'' 
have smaller feature sizes than the other gates and are used for fine tuning 
the confinement of each dot. The QPC-L and QPC-R gates are associated with 
the L- and R-gates (via the tips) to form the QPC detectors. Charging 
paths into the dots (shown by the ovals) from external reservoirs are shown 
by curved arrows, whereas the QPC currents are shown by straight arrows. 
Figure 1(b) shows a cross-sectional view of the layer structure in the 
$z$-direction. Our model involves four different layers of semiconductor 
materials (from top to bottom): a 50-\AA~thick $n$-type ($N_D=1.5\times10^{18}$ 
cm$^{-3}$) GaAs layer, a 650-\AA~thick $n$-type ($N_D=0.31\times10^{18}$ 
cm$^{-3}$) Al$_{0.27}$Ga$_{0.73}$As layer, a 200-\AA~thick 
undoped Al$_{0.27}$Ga$_{0.73}$As layer, and a 1610-nm thick $p$-type ($N_A=
1.0\times10^{15}$ cm$^{-3}$) GaAs layer. The 2DEG is formed at the 
interface between the undoped AlGaAs layer and the lightly $p$-type doped GaAs 
layer (900~\AA~below the top surface).

\begin{figure}[b]
\caption{\label{fig:fig1}
(a) Layout of the top gates (Light gray areas show the gate pattern for the LCQD 
and the QPC's; ovals show the dots; curved arrows show the 
possible charging current paths; and straight arrows show the QPC currents.). 
(b) Layers of the heterostructure (not to scale), 
after Elzerman {\it et al.}.\cite{Elzerman}}
\end{figure}

\section{Numerical Model}

 The electron density in the LCQD region is obtained by describing the charge carriers within the 
density functional theory (DFT) that incorporates many-body effects 
among particles.\cite{Jones} In order to take into account the spin 
dependence of the electron-electron interaction, the Kohn-Sham equations\cite{Kohn} 
for spin-up ($\uparrow$) and spin-down ($\downarrow$) are solved simultaneously:
$$
\begin{array}{c}
H^{\uparrow}\psi^{\uparrow}_i({\bf r})=\varepsilon^{\uparrow}_i\psi^{\uparrow}_i({\bf r}), \\
H^{\downarrow}\psi^{\downarrow}_i({\bf r})=\varepsilon^{\downarrow}_i\psi^{\downarrow}_i({\bf r}).
\end{array}
\eqno{(1)}
\label{KS}
$$
Here $\varepsilon^{\uparrow (\downarrow)}_i$ and  $\psi^{\uparrow (\downarrow)}_i$ 
are the corresponding eigenenergies and eigenfunctions of the Hamiltonian $H^{\uparrow (\downarrow)}$:
$$
H^{\uparrow (\downarrow)}=-\frac{\hbar^2}{2}\nabla\left[\frac{1}{m^*({\bf r})}\nabla\right]-q\phi({\bf r})\\
+\Delta E_c+\phi^{\uparrow (\downarrow)}_{xc}({n}),
\eqno{(2)}
$$
where $m^*({\bf r})$ is the position dependent effective mass. 
$\phi({\bf r})=\phi_{ext}+ \phi_{ion}+\phi_{H}$ is the electrostatic potential 
which consists three parts: $\phi_{ext}$ is the potential due to external gate biases, 
$\phi_{ion}$ is the potential resulting from ionized donors and acceptors, and $\phi_{H}$ is the 
Hartree potential accounting for repulsive electron-electron interactions. $\Delta E_c$ is 
the conduction-band offset between different materials, and 
$\phi^{\uparrow (\downarrow)}_{xc}({\bf r})$ is the exchange-correlation potential 
energy for spin-up ($\uparrow$) and spin-down ($\downarrow$) computed within 
the local spin density approximation (LSDA) according to Perdew and Wang's formulation.\cite{Perdew} 
Hence our approach is spin unrestricted by allowing for different orbitals with different spins. 

The electron density $n({\bf r})$ in the LCQD region is
$$
n({\bf r})=n^{\uparrow}({\bf r})+n^{\downarrow}({\bf r})=\\
\sum^{N_{\uparrow}}_{i=1}\left|\psi^{\uparrow}_i({\bf r})\right|^2+\sum^{N_{\downarrow}}_{i=1}\left|\psi^{\downarrow}_i({\bf r})\right|^2,
\eqno{(3)}
\label{ed}
$$
where $N_{\uparrow}+N_{\downarrow}= N$ is the total number of electrons in the dots.

The electrostatic potential $\phi({\bf r})$ is computed by solving Poisson's equation
$$
\nabla\left[\epsilon({\bf r})\nabla\phi ({\bf r})\right]=-\rho({\bf r}),
\eqno{(4)}
\label{P}
$$
where $\epsilon({\bf r})$ is the position-dependent permittivity and 
$\rho({\bf r})$ is the total charge density given by
$$
\rho ({\bf r})=q\left[N^+_D({\bf r})-N^-_A({\bf r})+p({\bf r})-n({\bf r})\right].
\eqno{(5)}
$$
Here $N^+_D ({\bf r})$ and $N^-_A ({\bf r})$ are the ionized donor and acceptor 
concentrations in the relevant device layers, $p({\bf r})$ is the hole concentration,
and $n({\bf r})$ is the total electron concentration given by Eq. (3) in the QD region, 
while outside this region the free electron charge is entirely determined by using the 
semi-classical Thomas-Fermi approximation.\cite{Mat2} 

We solve Kohn-Sham and Poisson equations self-consistently by finite element 
method.\cite{Mat1,Mat2} Zero normal electric field on lateral and bottom surfaces and 
Schottky barrier values on the top surface are imposed as boundary conditions for the 
solution of Poisson equation. Since the quantum dots are much smaller than the physical 
dimensions of the device, the wavefunctions actually vanish long before reaching 
the device boundaries. This allows us to embed a local region in the global mesh 
for solving the Kohn-Sham equations. This local region is chosen large enough to 
ensure vanishing wavefunctions on its boundaries. A non-uniform 3D grid of 141, 52 
and 71 mesh points in  the $x$-, $y$- and $z$-directions respectively, is used for 
solving Poisson equation, while $71\times45\times19$ grid points are used to discretize 
the local region where Kohn-Sham wavefunctions are evaluated.

Because the LCQD are weakly coupled to the external reservoirs, 
we assume that electrons in the dots are completely localized in that region. 
At equilibrium, and for a given bias, an integer number of electrons $N$ minimizes 
the total energy $E_{T}$ of the dots. 
In order to determine $N$, we use the Slater formula:\cite{Slater}
$$
E_T(N+1)-E_T(N)=\int_0^1\varepsilon_{LUO}(n)dn\\
\approx\varepsilon_{LUO}(1/2)-E_F,
\eqno{(6)}
$$
where $E_{T}(N+1)$, $E_{T}(N)$ are the total energies for $N+1$, $N$ electrons in the dots, 
and $\varepsilon_{LUO}(1/2)$ is the eigenenergy of ``the lowest unoccupied orbital'' 
with half occupancy. The sign change of the right-hand side of Eq. (6), as a function 
of the tuning gate voltage, determines the electron occupation in the LCQD. 
In our simulation, we use a variation of the above rule where charging occurs 
when $\varepsilon_{LUO}(1) - E_F = E_F - \varepsilon_{LUO}(0)$, which was justified in Ref. 12.

\begin{figure}[b]
\caption{\label{fig:fig2}
Conduction band edge profile in the LCQD-QPC structure (a) contour plot in the 
$xy$-plane at the 2DEG interface (The dashed rectangle shows the location of the dots.).
(b) along the $z$-direction with the inset showing the shape of the ground state 
wavefunction ($V_{PL}=V_{PR} = -0.15$ V, zero electrons in the LCQD).}
\end{figure}

\section{Results and Discussions}

Figures 2 shows the conduction band edge profiles in the $xy$-plane at the 2DEG 
interface (contour plot, Fig. 2(a)) and in the $z$-direction (Fig. 2(b)) 
under the condition $V_L=V_R=V_{QPC-L}=V_{QPC-R}=V_M= -0.585$ V, $V_T = -0.9$ V, 
$V_{PL}=V_{PR}= -0.15$ V (These voltages correspond to point A in Fig. 8.) and 
zero electrons in the dots. The Fermi level is set at zero throughout the device 
at the temperature $T = 4$ K. The LCQD region and the QPC region with low 
equipotential-line density are clearly visible in Fig. 2(a). The outer energy 
barrier for the LCQD is $\sim110$ meV whereas the energy barrier between the dots 
is $\sim9$ meV. A large negative T gate bias is used to prevent the wavefunctions 
from leaking into the external reservoirs, which clearly defines the LCQD region. 
The confinement along the $z$-direction is achieved by a quasi-triangular shaped 
well shown in Fig. 2(b), for which the relaxation of the potential to zero-field 
is not shown at the far end (substrate) of the device. Due to the strong 
confinement in the triangular well, only the ground state along the $z$-direction 
is occupied (the shape of the ground state wavefunction along the $z$-direction 
is shown in the 
inset in Fig. 2(b).). Under the above condition, the wavefunction contour plots 
in the $xy$-plane at the 2DEG interface are shown in ascending energies 
for the first eight spin-up 
($\uparrow$) eigenstates in Fig. 3. A similar set of wavefunctions is obtained 
for the spin-down ($\downarrow$) eigenstates (not shown). They are similar to 
orbitals observed in diatomic molecules: the two columns represent the familiar 
bonding and anti-bonding state pairs. Notice that the shape of the wavefunctions 
reflects the shape of the confinement seen in the local minima of the conduction 
band edge in Fig. 2(a).

\begin{figure}[t]
\caption{\label{fig:fig3}
Contour plot of the first eight spin-up ($\uparrow$) eigenstates in ascending energies in the 
$xy$-plane at the 2DEG interface with zero electrons in the LCQD ($V_{PL} = V_{PR} =
-0.15$ V). The $xy$-coordinates are given for the lower left wavefunction, which is 
a zoom-in region corresponding to the dashed rectangular region in Fig. 2(a); all 
the other wavefunction contour plots in this paper are on the same scale.}
\end{figure}

\begin{figure}[t]
\caption{\label{fig:fig4}
(a) Eigenenergy spectrum (spin-up ($\uparrow$) states) as a function of the right 
plunger gate bias (solid lines: right dot; dashed lines: left dot). 
$\alpha$, $\beta$ and $\gamma$ are three ``anti-crossing'' points. 
(b) Variation of the conduction band edge in the constriction of the left and 
right QPC's as a function of the right plunger gate bias from point A to B in Fig. 8
($V_{PL}$ is fixed to $-0.15$ V; the vertical axis of Fig. 4(b) is shifted up by $0.0201$ eV.).}
\end{figure}

In Figure 4(a), we show the variation of the first eight spin-up ($\uparrow$)
eigenenergies when the 
plunger gate bias configuration is changed from the values 
$V_{PL}=V_{PR}= -0.15$ V to the new values $V_{PL}= -0.15$ V, $V_{PR}= -0.06$ V. 
The first eight eigenenergies are separated into two groups, one for the right 
dot (solid lines) and one for the left dot (dashed lines), which are lowered 
simultaneously as the right plunger gate bias increases. However, the 
eigenenergies of the right dot decrease more rapidly than those of the left dot 
because of the proximity of the former to the varying plunger. At 
$V_{PR} =-0.074$ V, the charging of the first electron (spin-up ($\uparrow$)) 
occurs in the right dot, 
which is indicated by a discontinuity of $8.2\times10^{-4}$ eV in the 
variation of the ground state energy level with respect to the right plunger 
gate bias. At the same gate bias, we also observe a jump of the conduction band 
edge in the constriction of the two QPC's, {\it i.e.}, $2.6\times10^{-6}$ eV for the 
left QPC and $5.4\times10^{-6}$ eV for the right QPC (see Fig. 4(b), where the 
vertical axis is shifted up by 0.0201 eV for clarity). The up-shift of the 
conduction band edge in the QPC constriction results from the Coulomb interaction 
between the electrons in the LCQD and electrons in the QPC's, which reduces the 
total charge number in the conduction channel and leads to a discontinuity in the 
QPC current observed in experiments.\cite{Elzerman} Obviously, the right QPC 
is more sensitive to the single-electron charging because of its proximity with 
the right dot. On the stability diagram (Fig. 8), this transition is represented 
by the vertical A to B line with the diamond indicating the charging point for 
the first electron.

From the eigenenergies variation vs. $V_{PR}$ diagram (Fig. 4(a)), we also 
observe three ``anti-crossing'' points between the two different sets of 
eigenenergy levels, each arising from the distinct QD's as mentioned above and 
indicated by arrows in Fig. 4(a), {\it i.e.}, ($\alpha$) at $V_{PR}\sim-0.09$ V, 
between the 3rd and 4th excited states; ($\beta$) at $V_{PR}\sim-0.07$ V, 
between the 3rd and 4th excited states; and ($\gamma$) at $V_{PR}\sim-0.11$ V, 
between the 5th and 6th excited states. The behavior of the system near the 
``anti-crossing'' points can be further illustrated by examining the evolution 
of the wavefunctions for the ``anti-crossing'' levels. ``Interchange'' of the 
wavefunctions is clearly observed before and after these points. In Fig. 5, 
contour plots of the wavefunctions in the $xy$-plane at the 2DEG interface are 
shown for the three ``anti-crossing'' points: the 3rd and 4th excited states labeled
$\alpha_1$, $\alpha_2$ at $V_{PR} = -0.10$ V and $\alpha'_1$, $\alpha'_2$ at 
$V_{PR}= -0.08$ V, respectively; the 3rd and 4th excited states labeled 
$\beta_1$, $\beta_2$ at $V_{PR} = -0.074$ V and $\beta'_1$, $\beta'_2$ at 
$V_{PR}= -0.06$ V; and the 5th and 6th excited states labeled $\gamma_1$, 
$\gamma_2$ at $V_{PR} = -0.12$ V and $\gamma'_1$, $\gamma'_2$ at $V_{PR}= -0.10$ V.

\begin{figure}[b]
\caption{\label{fig:fig5}
Wavefunction (for spin-up ($\uparrow$) states) ``interchanges'' at the ``anti-crossing'' points 
corresponding to (a) point $\alpha$, (b) point $\beta$ and (c) point $\gamma$ 
in Fig. 4(a).}
\end{figure}

The detection of single-electron charging events can also be carried out for the 
B to C transition in Fig. 8, in which the right plunger gate bias $V_{PR}$ is 
fixed to be $-0.06$ V while the left plunger gate bias is changed from 
$V_{PL} = -0.15$ V to $V_{PL} = -0.06$ V. The variation of the spin-up ($\uparrow$)
eigenenergies with respect to the 
Fermi level and the conduction band edge in the constriction of the two QPC's are 
shown in Figs. 6(a) and (b), respectively. In this case, the transition of the 
charging state is from one electron in the right dot to two electrons, one in each 
dot occupying an individual $1S$-like orbital,\cite{Nagaraja} and occurs when the 
left plunger gate bias $V_{PL}$ is at $-0.097$ V. The charging of the second 
electron into the system is localized in the left dot and is indicated by the jump of 
the first excited state energy level. Note that in this case, the variation of 
eigenenergies in the left QD (dashed lines) is larger than those in the right dot 
(solid lines). In our LSDA 
approach, the second electron has the same spin (spin-up ($\uparrow$)) as the first 
one as they are 
uncorrelated by the height of the coupling barrier. The corresponding jump of 
the conduction band edge is $5.6\times10^{-6}$ eV for the left QPC and 
$2.8\times10^{-6}$ eV for the right one. The left QPC is more sensitive to the
second electron charging because it occurs in the left dot. Fig. 7 shows the 
spin-up ($\uparrow$) wavefunctions in ascending energies (from top to bottom) 
after the charging of the second electron for two bias conditions, 
1) $V_{PL} = -0.08$ V and 2) $V_{PL} = -0.06$ V (for both cases, $V_{PR} =-0.06$ V). 
It demonstrates the evolution of the wavefunctions from an asymmetric 
configuration with different eigenenergy levels for two electrons in the system 
to a symmetric one where eigenenergy levels are fully degenerate.

\begin{figure}[t]
\caption{\label{fig:fig6}
(a) Eigenenergy spectrum (spin-up ($\uparrow$) states) (solid lines: right dot; 
dashed lines: left dot) and (b) variation of the 
conduction band edge in the constriction of the left and right QPC's as a 
function of the left plunger gate bias from point B to C in Fig. 8
($V_{PR}$ is fixed to $-0.06$ V; the vertical axis of Fig. 6(b) is shifted up 
by $0.0201$ eV.).}
\end{figure}

\begin{figure}[htb]
\caption{\label{fig:fig7}
Evolution of the first eight spin-up ($\uparrow$) wavefunctions from bias condition 
1) $V_{PL} = -0.08$ V to 2) $V_{PL} =-0.06$ V (for both cases, $V_{PR} = -0.06$ V).}
\end{figure}

Following the same procedure as described above, we can find another charging 
path for the first electron charging, {\it i.e.}, from point E to F in Fig. 8, 
and for the charging from one to two electrons, F to G in Fig. 8, for distinct 
stable charge regimes of electrons in the two dots. On the path E to F ($V_{PL}$ 
is fixed to $-0.125$ V, $V_{PR}$ is changed from $-0.125$ V to $-0.07$ V), 
charging happens for the first electron (spin-up ($\uparrow$)) in the right dot 
at $V_{PR} = -0.082$ V; 
on the path F to G ($V_{PR}$ is fixed to $-0.07$ V, $V_{PL}$ is changed from 
$-0.125$ V to $-0.07$ V), charging happens for the second electron (spin-up ($\uparrow$)) 
in the left dot at $V_{PL} = -0.092$ V.

We can further interchange the plunger gate biases and obtain different 
transitions, {\it i.e.}, from A to D to C and E to H to G, as shown in Fig. 8 
to realize closed cycles of charging and discharging paths. These two closed paths 
(dashed and dotted lines) are shown in Fig. 8. Each corner 
of the two squares is in a different stable charge state with numbers in the 
parentheses showing the electron number in the left and right dots, respectively, 
{\it e.g.}, (0,1) means zero electron in the left dot and one in the right dot. On each 
path, we record the charging points (diamonds in Fig. 8) and make linear 
extrapolations between the two charging points on each two parallel paths, which 
leads to four lines crossing at two points (circles in Fig. 8).

The two crossing points are linked afterwards. Now, five segments (solid lines in 
Fig. 8) separate the diagram into four regions to define the stability diagram for 
the LCQD system in the few-electron charging regime. Each region, separated by the 
solid lines, indicates a stable charge configuration assumed by the LCQD under a 
particular range of plunger gate biases. More interesting are the two crossing points 
(circles), called the double-triple point,\cite{Wiel} occurring at 
$V_{PL}=V_{PR} = -0.0924$ V for the three charging states (0,0), (0,1) and (1,0) 
and at $V_{PL}=V_{PR}= -0.0847$ V for (0,1), (1,0) and (1,1) states. We then 
determine the voltage range of the right plunger that spans the distance between 
the double-triple point to be $\Delta V_{PR}=7.7$ mV, which is comparable to the 
experimental result $\sim7.4$ mV.\cite{Elzerman} 

\begin{figure}[t]
\caption{\label{fig:fig8}
Stability diagram for the first two charging electrons characterizing 
the double-triple point (shown by circles).}
\end{figure}

Finally, from the charging diagrams in the few-electron regime we extract the addition 
energy for the second electron charging in the right dot: we determine the 
$V_{PR}$-voltage interval on the stability diagram for the (0,1) configuration 
({\it i.e.}, between the (0,0) configuration and the (0,2) configuration in the 
singlet state) to be $0.1$ V, which is in excellent agreement with the experimental 
result $\sim0.1$ V.\cite{Elzerman} By linear projection of this $V_{PR}$ interval to the 
energy scale,\cite{Mat2} we obtain then the addition energy for charging the second 
electron, which is $2.5$ meV. By comparing this value to the experimental result of 
$3.7$ meV,\cite{Elzerman} we attribute the difference to the fact that our simulation 
is performed on a coupled-dot system, while the experimental result is obtained by 
grounding one of the dots where the confinement is stronger in an individual dot 
compared to our simulation case.

\section{Conclusion}

We performed numerical simulations of the electrostatic and quantum-mechanical 
characteristics of a novel laterally-coupled quantum dot circuit with integrated 
quantum point contact read-out. We were able to reproduce detailed single-electron 
charging behavior of the elementary quantum circuit and demonstrate the 
functionality of the QPC's as single-electron charging detectors. In particular, 
we obtained excellent agreement with the experiment for the voltage range of the 
extension of the double-triple point at the (0,0) to (1,1) transition and the 
addition energy for single-electron charging in the dots, which 
validates our quantum device modeling approach for simulating efficiently 
nanoscale qubit circuits.

\begin{acknowledgments}
This work is supported by DARPA QUIST program through ARO Grant DAAD 19-01-1-0659.
The authors thank D. Melnikov for constructive discussions. L.-X. Zhang thanks 
R. Ravishankar, S. Barraza-Lopez for their technical assistance.
\end{acknowledgments}

\end{document}